\documentclass[epj]{webofc}
\usepackage[varg]{txfonts}   
\usepackage{amsmath}
\newcommand{\beq}{\begin{equation}}
\newcommand{\eeq}{\end{equation}}
\newcommand{\ba}{\begin{eqnarray}}
\newcommand{\ea}{\end{eqnarray}}

\newcommand{\bm}{\boldsymbol} 

\woctitle{6th International conference on Physics Opportunities at an ElecTron-Ion Collider (POETIC VI)}
\begin{document}
\title{Gluon TMD studies at EIC}
%
%

\author{Dani\"el Boer\inst{1}\fnsep\thanks{\email{d.boer@rug.nl}} 
}

\institute{Van Swinderen Institute for Particle Physics and Gravity, University of Groningen\\ 
Nijenborgh 4, NL-9747 AG Groningen, The Netherlands}

\abstract{%
A high-energy Electron-Ion Collider (EIC) would offer a most promising tool to study in detail the transverse 
momentum distributions of gluons inside hadrons. This applies to unpolarized as well as linearly polarized 
gluons inside unpolarized protons, and to left-right asymmetric distributions of gluons inside transversely 
polarized protons, the so-called gluon Sivers effect. The inherent process dependence of these distributions 
can be studied by comparing to similar, but often complementary observables at LHC. 
}
\maketitle
\section{Introduction}
\label{intro}
Transverse momentum dependent parton distributions (TMDs) are currently under active investigation, 
both theoretically and experimentally. Typical TMD processes are semi-inclusive Deep Inelastic Scattering (SIDIS) 
or the Drell-Yan process. The SIDIS process ($e\, p \to e' \, h\, X$) is sensitive to the transverse momentum of quarks, 
while for instance $D$-meson pair production ($e\, p \to e' D\, \overline{D}\, X$) is sensitive to the transverse momentum of gluons 
in the back-to-back correlation limit. For studies of the gluon TMDs, higher energy ($\!\sqrt{s}$) or smaller 
$x$ is required. A high-energy Electron-Ion Collider (EIC) can offer clean probes of 
the distributions of unpolarized and linearly polarized gluons inside unpolarized protons, and of the gluon Sivers effect for 
transversely polarized protons. These distributions and what we can learn about them at an EIC will be reviewed here, with emphasis on the most promising observables, the process dependence, and the expected small-$x$ behavior of the distributions. 

Describing the transverse momentum of partons in a process is not just a matter of adding a transverse momentum dependence in collinear distributions, i.e.\ $f_1(x) \to  f_1(x,k_T^2)$, that appear in collinear factorization expressions. Rather one has to deal with TMD factorization, in which new factors and new distributions appear, such as the Sivers effect TMD that describes a correlation between the transverse momentum and the proton spin.  

For gluons there are eight leading twist TMDs \cite{Mulders:2000sh} that parametrize the gluon correlator
\beq
\Gamma_g^{\,\mu\nu\, [{\cal U},{\cal U}^\prime]}(x,k_T) \equiv	\int \frac{d(\xi\cdot P)\, d^2 \xi_T}{(x P\cdot n)^2 (2\pi)^3} e^{i ( xP + k_T) \cdot \xi} \langle P| {\rm Tr}_c \Big[ F^{n\nu}(0)\, {\cal U}_{[0,\xi]}\, F^{n\mu}(\xi)\, {\cal U}_{[\xi,0]}^\prime \Big] |P\rangle_{\xi \cdot P^\prime = 0}. \label{GammaUU} 
\eeq
The dependence on the gauge links ${\cal U}$ and ${\cal U}^\prime$ will be discussed later on. 
For unpolarized hadrons the correlator $\Gamma_g$ is parametrized by two gluon TMDs \cite{Mulders:2000sh} (here $\boldsymbol{k}_T^2= -k_T^2$):
\beq
\Gamma_g^{\,\mu\nu}(x,\boldsymbol{k}_T )
= -\frac{1}{2x}\,\bigg \{g_T^{\mu\nu}\,f_1^{\, g}
-\bigg(\frac{k_T^\mu k_T^\nu}{M^2}\,
{+}\,g_T^{\mu\nu}\frac{\boldsymbol{k}_T^2}{2M^2}\bigg)
\;h_1^{\perp\,g} \bigg \}.
\eeq
The unpolarized gluon TMD $f_1^{\, g}$ and linearly polarized gluon TMD $h_1^{\perp\, g}$ are both functions of $x$ and $\boldsymbol{k}_T^2$.  
Nonzero $h_1^{\perp\, g}$ requires nonzero transverse momentum and stems from an interference between $\pm 1$ gluon helicities. For positive $h_1^{\perp\, g}$ the gluon polarization $\boldsymbol{\epsilon}_T$ is distributed around $\boldsymbol{k}_T$ with a $\cos 2\phi$ distribution ($\phi = \angle (\boldsymbol{k}_T,\boldsymbol{\epsilon}_T)$).  Linear gluon polarization modifies among others the transverse momentum distribution of Higgs production (perturbatively at NNLO \cite{Catani:2010pd,Wang:2012xs} and nonperturbatively at LO in pQCD \cite{Sun:2011iw,Boer:2011kf}), which can be studied at LHC.
In TMD factorization the cross section takes the form \cite{Boer:2011kf}
\beq
\frac{E\,d\sigma^{p p \to H X}}{d^{3}\vec{q}}\Big|_{q_{T}\ll m_{H}} \propto \left(\mathcal{C}\left[f_{1}^{\,g}\, f_{1}^{\,g}\right]+\mathcal{C}\left[w_{H}\, h_{1}^{\perp \, g}\, h_{1}^{\perp \, g}\right]\right)\,+\mathcal{O}\left(\frac{q_{T}}{m_{H}}\right),
\eeq
where ${\cal C}$ denotes a convolution of TMDs and $w_{H}=\left((\bm{k}_{1T}\cdot\bm{k}_{2T})^{2}-\frac{1}{2}\bm{k}_{1T}^{2} \bm{k}_{2T}^{2}\right)/2M^4$. Including resummation of large logarithms the contribution of linearly polarized gluons relative to unpolarized gluons is given by (for explanations cf.\  \cite{Boer:2011kf}):
\beq
{\cal R}(Q_T) \equiv \frac{\mathcal{C}[w_H\,h_1^{\perp\, g}\,h_1^{\perp\, g}]}{\mathcal{C}[f^{\, g}_1\,f_1^{\, g}]} =  \frac{\int d^2 \bm{b} \, e^{i \bm{b} \cdot \bm{q}_T^{}} e^{-S_A(b_*,Q) - S_{NP}(b,Q)}
\; \widetilde{h}_1^{\perp\, g}(x_{A},b_*^2; \mu_{b_*}) \; \widetilde{h}_1^{\perp\, g}(x_{B},b_*^2; \mu_{b_*})}{
\int d^2 \bm{b} \, e^{i \bm{b} \cdot \bm{q}_T^{}} \, e^{-S_A(b_*,Q)- S_{NP}(b,Q)} \widetilde{f}_1^{\, g}(x_{A},b_*^2; \mu_{b_*}) \, 
\widetilde{f}_1^{\, g}(x_{B},b_*^2; \mu_{b_*})}, \label{calRQT}
\eeq
where $\widetilde{f}_1^{\, g}$ denotes the Fourier transform of $f_1^{\, g}$ and
\beq
\widetilde{h}_1^{\perp\,g}(x,b^2) \equiv \int d^2\bm{k}_T^{}\; \frac{(\bm{b}\!\cdot \!
\bm{k}_T^{})^2 - \frac{1}{2}\bm{b}^{2} \bm{k}_T^{2}}{b^2 M^2}
\; e^{-i \bm{b} \cdot \bm{k}_T^{}}\; h_1^{\perp\, g}(x,k_T^2) =  -\pi \int dk_T^2
\frac{k_T^2}{2M^2} J_2(bk_T) h_1^{\perp\, g}(x,k_T^2).
\eeq
The integrand in $b$ space has been split into a calculable perturbative part and a nonperturbative (NP) part that should be obtained from fits to data. Although the nonperturbative Sudakov factor $S_{NP}$ for $g g \to H$ is unknown, at the Higgs scale it does not matter too much. What matters most is the small-$b$ part of the TMDs, which is perturbatively calculable \cite{Nadolsky:2007ba,Catani:2010pd,Sun:2011iw}: $\widetilde{f}_1^{\, g}(x,b^2; \mu_b) = f_{g/P}(x; \mu_b) +  {\cal O} (\alpha_s)$, while 
\beq
\widetilde{h}_1^{\perp\, g}(x,b^2; \mu_b) =  \frac{\alpha_s(\mu_b) C_A}{2\pi}
\int_x^1 \frac{d\hat x}{\hat x} \left(\frac{\hat x}{x}-1\right) f_{g/P}(\hat x; \mu_b) 
+ {\cal O} (\alpha_s^2). \label{tailh1perp}
\eeq
Note that the perturbative tail of $h_1^{\perp\, g}$ is driven by the unpolarized collinear gluon distribution $f_{g/P}(x;\mu)$. 

In \cite{Boer:2014tka} and \cite{Echevarria:2015uaa}  the above expressions were studied numerically (cf.\ \cite{Boer:2015uqa}
for the ranges of the predictions). The conclusion from those studies is that ${\cal R}(Q_T)$ is on the order of 2-5\% in Higgs production at low $Q_T$. This probably means that the extraction of $h_1^{\perp\, g}$ from Higgs production will be too challenging. In \cite{Boer:2014tka} and \cite{Echevarria:2015uaa} also heavy (pseudo-)scalar $C=+$ quarkonium production, $p\,p \to [\overline{Q}Q]\, X$, has been studied. Much larger effects from linear gluon polarization are possible in this case, but there are very large uncertainties (cf.\ \cite{Boer:2015uqa}). 
It is much more sensitive to the unknown NP part than Higgs production. From this perspective the heavier bottomonium states are probably best to consider.  
Employing the color singlet model \cite{Baier:1983va} and LO NRQCD results \cite{Bodwin:1994jh,Bodwin:2005hm}, the differential cross sections for $\eta_b$, $\chi_{b 0}$ and $\chi_{b 2}$ production have been obtained in \cite{Boer:2012bt}. By forming ratios of ratios, in which the hadronic uncertainties cancel, 
it becomes in principle possible to probe ${\cal R}(Q_T)$ directly: 
\beq
\frac{\sigma (\chi_{b 2})}{\sigma(\chi_{b 0})} \frac{d\sigma(\chi_{b 0})/d^2 \bm{q}_T}{d\sigma (\chi_{b 2})/d^2 \bm{q}_T}  \approx  1 + {\cal R}(Q_T), \qquad
\frac{\sigma (\chi_{b 0})}{\sigma(\eta_b)} \frac{d\sigma(\eta_b)/d^2 \bm{q}_T}{d\sigma (\chi_{b 0})/d^2 \bm{q}_T} \approx  \frac{1 - {\cal R}(Q_T)}{1 + {\cal R}(Q_T)}.\label{ratioofratios}
\eeq
These are color singlet model expressions, which may be justified
for $C=+$ bottomonium states from NRQCD considerations \cite{Hagler:2000dd,Bodwin:2005hm} and by several numerical studies of color octet contributions \cite{Hagler:2000dd,Maltoni:2004hv,Lansberg:2012kf}.
TMD factorization for the $p$-wave states $\chi_{bJ}$ has been called into question though \cite{Ma:2014oha}. Consistency between the experimental results for ${\cal R}(Q_T)$ from (\ref{ratioofratios}), e.g.\ at LHCb, can be used to assess the possible factorization breaking contributions. Because of the small energy scale differences ($m_{\eta_b} =9.4$ GeV, $m_{\chi_{b0}}=9.9$ GeV, $m_{\chi_{b2}} =10.3$ GeV), evolution effects should be negligible in this comparison. 

\section{EIC probes}
At EIC $h_1^{\perp\, g}$ can be probed in open charm and bottom quark pair electro-production, $e\, p \to e' \, Q\, \overline Q\, X$, where $Q$ and $\overline Q$ are almost back-to-back in the transverse plane. Unlike Higgs production one needs to study angular distributions now, e.g.\ a $\cos 2\phi$ asymmetry where $\phi = \phi_T - \phi_\perp$ and $\phi_{T/\perp}$ are the angles of 
$K_\perp^Q \pm K_\perp^{\overline Q}$ \cite{Boer:2010zf}, under the restriction $q_T \equiv (K_\perp^Q + K_\perp^{\overline Q}) \ll K_\perp\equiv (K_\perp^Q - K_\perp^{\overline Q})/2$. In the asymmetry expression, $h_1^{\perp\, g}$ appears by itself, as opposed to in a product of two. Therefore, larger effects are expected and the sign of  $h_1^{\perp\, g}$ can be determined. 
The asymmetry depends on $Q^2, K_\perp^2$ and $M_Q^2$, but the maximum of the asymmetry is to a large extent independent of these scales, and around 15\% \cite{Pisano:2013cya}. There are also angular asymmetries w.r.t.\ the lepton scattering plane that probe $h_1^{\perp\, g}$. These are mostly relevant at smaller $|K_\perp|$ \cite{Pisano:2013cya}. Dijet DIS, $e\, p \to e' \, {\rm jet} \, {\rm jet}\, X$, is similar except that also quark TMDs enter. 
The analogous processes $p\, p \to Q\, \overline{Q}\, X$ and $p\, p \to {\rm jet} \, {\rm jet}\, X$ at RHIC or LHC are not expected to be TMD factorizing \cite{Rogers:2010dm}.

At an EIC one also can consider transversely polarized protons, where $e\, p^\uparrow \to e' \, Q\, \overline Q\, X$ is a very promising process for probing the gluon Sivers effect. For a review of the status and prospects of the gluon Sivers distribution, cf.\ \cite{boer:2015ika}, and for specific model studies, cf.\ \cite{Boer:2011fh}. There are also suggestions to measure the gluon Sivers effect in proton-proton/ion collisions (RHIC, AFTER@LHC), in processes for which TMD factorization may hold: $p^\uparrow\,p \to \gamma \,{\rm jet}\, X$ \cite{Schmidt:2005gv,Bacchetta:2007sz}, $p^\uparrow\,p \to J/\psi \, \gamma \, X$ \cite{Dunnen:2014eta}, $p^\uparrow\,p \to J/\psi \,J/\psi \, X$ \cite{Lansberg:2014myg,Lansberg:2015lva}. According to~\cite{Dunnen:2014eta,Lansberg:2014myg}, the color singlet contribution to a large extent dominates over the color octet one in the $J/\psi$ production processes.  

Such gluon Sivers effect measurements in $p\, p$ collisions are complementary to EIC studies, because TMDs are actually process dependent, as will be discussed next. Although this process dependence can be calculated, not all Sivers functions 
from all processes can be related to each other!

\section{Process dependence}
It has been realized that TMDs in general are not universal \cite{Brodsky:2002cx,Collins:2002kn,Belitsky:2002sm}. Gluon rescattering corrections can be summed into path-ordered exponentials ${\cal U}_{{\cal C}}$ in TMD correlators \cite{Efremov:1978xm}, where the gauge link or Wilson line is along a path ${\cal C}$. The path in this Wilson line depends on whether color charges are coming from the initial state or going into the final state \cite{Collins:1983pk,Boer:1999si,Collins:2002kn,Brodsky:2002rv,Belitsky:2002sm,Boer:2003cm}. Surprisingly, it has turned out that in certain cases the shape of the Wilson lines affects observables, such as the Sivers effect asymmetries. In SIDIS the quark TMD correlator has a future pointing staple-like Wilson line 
arising from final state interactions (FSI), referred to as a $+$ link. In the Drell-Yan (DY) process it is past pointing from initial state interactions (ISI), a $-$ link. The quark Sivers functions with $+$ and $-$ links are related by parity and time reversal invariance by an overall minus sign:  
$f_{1T}^{\perp [{\rm SIDIS}]} = - f_{1T}^{\perp [{\rm DY}]}$ \cite{Collins:2002kn}. In general, the more hadrons observed in a process, the more complicated the resulting Wilson lines and the possible relations among TMDs of various processes \cite{Bomhof:2004aw,Buffing:2012sz,Buffing:2013dxa}. Wilson lines may even become entangled or trapped, leading to factorization breaking \cite{Collins:2007nk,Rogers:2010dm}.

The processes that allow access to the linearly polarized gluon distribution and the gluon Sivers distribution
depend on two gauge links as in Eq.\ (\ref{GammaUU}). The subprocess $\gamma^*\, g \to Q \, \overline{Q}$ for 
$e\, p \to e' \, Q\, \overline Q\, X$ probes a gluon correlator with two $+$ links, i.e.\ both are future pointing. 
In the kinematic regime where gluons in one proton in $p\,p \to \gamma \,{\rm jet}\, X$ dominate, 
one effectively selects the subprocess $q \, g \to \gamma \, q$. The latter subprocess probes a gluon correlator with a $+$ and $-$ link (future and past pointing), enclosing an area. As a consequence, these two processes probe two distinct, independent gluon Sivers functions. They correspond to antisymmetric ($f_{abc}$) and symmetric ($d_{abc}$) color structures as discussed in \cite{Buffing:2013kca}. 

At LHC $gg \to H$ and $gg \to [\overline{Q}Q]$ both probe a gluon correlator with two $-$ links. As $h_1^{\perp \, g \, [+,+]}= h_1^{\perp \, g \, [-,-]}$ (and $h_1^{\perp \, g \, [+,-]}= h_1^{\perp \, g \, [-,+]}$), one concludes that EIC and LHC can probe the same $h_1^{\perp\, g}$ function. But e.g.\ $gg \to Hg$ probes a more complicated link structure. On the other hand, for the gluon Sivers function it holds that $f_{1T}^{\perp \, g \, [+,+]}= - f_{1T}^{\perp \, g \, [-,-]}$ and $f_{1T}^{\perp \, g \, [+,-]}= - f_{1T}^{\perp \, g \, [-,+]}$. One thus concludes that the proposed gluon Sivers TMD studies at EIC and at RHIC or AFTER@LHC are complementary \cite{boer:2015ika}. 

This TMD nonuniversality is not just a polarization issue. It was first realized in a small-$x$ context that this process dependence also applies to the unpolarized gluon TMD $f_1^{\, g}$ \cite{Dominguez:2011wm}. 

\section{Small-x: a tale of two gluon distributions}
At small $x$ (and large $N_c$) there are two unpolarized gluon distributions that matter \cite{Dominguez:2011wm}, 
the gluon correlator with two $+$ links (for which $f_1^{\,g \, [+,+]}= f_1^{\,g \, [-,-]}$) and the one with a $+$ and a $-$ link
(for which $f_1^{\,g \, [+,-]}= f_1^{\,g \, [-,+]}$). In \cite{Dominguez:2011wm} these were denoted by $G^{(1)}$ and $G^{(2)}$, respectively. 
At small $x$ they correspond to the Weizs\"acker-Williams (WW) and dipole (DP) distributions, which are in general different. The fact that there are two distinct but equally valid definitions for the gluon distribution was noted first in ``A tale of two gluon distributions'' by Kharzeev, Kovchegov \& Tuchin (KKT) in \cite{Kharzeev:2003wz}, where the authors say that they ``cannot offer any simple physical explanation of this paradox''. The explanation turns out to be the process dependence of the gluon distribution, in other words, its sensitivity to the ISI/FSI in a process. Here it is not so much the direction, but rather whether a process is only sensitive to {\it either} ISI or FSI or to {\it both} ISI and FSI. The difference between the WW and DP distributions would disappear without ISI/FSI.   
In the MV model considered by KKT, one may not notice the origin for the difference, because the two gluon distributions become related: $xG^{(2)}(x,q_\perp) \stackrel{MV}{\propto} q_\perp^2 \nabla_{q_\perp}^2 x G^{(1)}(x,q_\perp)$ \cite{Kharzeev:2003wz, Dominguez:2011wm}.
For instance, the process $\gamma \, A \to Q \, \overline{Q} \, X$ has been studied in \cite{Gelis:2001da} in the MV model, where the cross section $d\sigma_T/dydk_\perp$ is expressed in terms of $C(k_\perp)= \int d^2x_\perp e^{ik_\perp \cdot x_\perp} \langle U(0)U^\dagger(x_\perp)\rangle \sim G^{(2)}$ which is the DP distribution, whereas the process rather probes the WW distribution $G^{(1)}$ ($[+,+]$). 

Different processes probe $G^{(1)}$ or $G^{(2)}$ or a mixture, as listed in Table \ref{tab-1}.
\begin{table}
\centering
\caption{List of processes that probe the WW and/or DP unpolarized gluon TMD at small $x$ \cite{Dominguez:2011wm}.}
\label{tab-1}       
\begin{tabular}{|l|c|c|c|c|c|c|}
\hline
\hline
{}& DIS \& DY & SIDIS & $p\, A \to h\, X$ & $pA\to \gamma\, {\rm jet}\, X$ & Dijet in DIS & Dijet in $pA$ \\\hline
$f_1^{\,g \, [+,+]}$ (WW) & $\times$ & $\times$ & $\times$ & $\times$ & $\surd$ & $\surd$ \\\hline
$ f_1^{\,g\, [+,-]}$ (DP) & $\surd$ & $\surd$ & $\surd$ & $\surd$ & $\times$ & $\surd$ \\\hline\hline
\end{tabular}
\end{table}
For dijet production in $p\,A$ collisions, the result requires large $N_c$, otherwise (four) additional functions appear (cf.\ \cite{Kotko:2015ura}).  

This process dependence of TMDs implies that also their $p_T$-widths are process dependent, and as a consequence, it gives an 
additional process dependence to $p_T$-broadening \cite{Boer:2015kxa}.

The WW and DP $h_1^{\perp \, g}$ distributions will be different too. Within the MV model \cite{Metz:2011wb} the DP $h_1^{\perp \, g}$ distribution is found to be maximal for all transverse momenta, while the WW $h_1^{\perp \, g}$ distribution is maximal only at large $k_T$ ($\gg Q_s$) and suppressed w.r.t.\ $f_1^{\, g}$ in the saturation region ($k_T \ll Q_s$). Maximal (positive) linear polarization also arises in the small-$x$ ``$k_T$-factorization'' approach \cite{Catani:1990eg}:
\beq
 \Gamma_g^{\,\mu\nu}(x,\bm{k}_T )_{\rm max\ pol}= \frac{1}{x}\,\frac{k_T^\mu k_T^\nu}{\bm{k}_T^2}\,f_1^{\, g}(x,\bm{k}_T^2).
\eeq
Finally, the perturbative tail of $h_1^{\perp \, g}$ in Eq.\ (\ref{tailh1perp}) has a $1/x$ growth, which keeps up with $f_1^{\, g}$ towards small $x$. Clearly there is no theoretical reason why $h_1^{\perp \, g}$ should be small, especially at small $x$. In analogy to $f_1^{\, g}$, in Table \ref{tab-2} we list processes where the WW and DP $h_1^{\perp \, g}$ distributions enter (or not). 
\begin{table}
\centering
\caption{List of processes that probe the WW and/or DP linearly polarized gluon TMD at small $x$.}
\label{tab-2}       
\begin{tabular}{|l|c|c|c|c|c|c|}
\hline
\hline
{}& DIS \& DY & SIDIS & $p\, A \to h\, X$ & $pA\to \gamma\, {\rm jet}\, X$ & Dijet in DIS & Dijet in $pA$ \\\hline
$h_1^{\perp \, g \, [+,+]}$ (WW) & $\times$ & $\times$ & $\times$ & $\times$ & $\surd$ & $\surd$ \\\hline
$ h_1^{\perp \, g \, [+,-]}$ (DP) & $\times$ & $\times$ & $\times$ & $\times$ & $\times$ & $\surd$ \\\hline\hline
\end{tabular}
\end{table}
It turns out that the processes DIS, DY, SIDIS, hadron and $\gamma+{\rm jet}$ production in $pA$ collisions do not probe $h_1^{\perp \, g}$ in leading power \cite{Boer:2009nc}. Dijet production in $ep$ and $eA$ collisions at small $x$ probes the WW distribution. Since there are different expectations inside and outside the saturation region, 
it would thus be very interesting to study $h_1^{\perp \, g}$ (via the $\cos 2 \phi$ asymmetries discussed earlier) in dijet DIS at a high-energy EIC. The relevant expressions for general $x$ can be found in \cite{Pisano:2013cya} and small-$x$ expressions in \cite{Metz:2011wb,Dumitru:2015gaa}. As said, these expressions involve only the WW-type distributions (at any $N_c$). 
In contrast, dijet and open heavy quark pair production in $p\,p$ and $p\,A$ collisions suffer from factorization breaking \cite{Rogers:2010dm}. Although at small $x$ the factorization breaking contributions may become suppressed, effectively restoring TMD factorization \cite{Chirilli:2011km,Kotko:2015ura}, still a combination of six distinct distributions is probed, complicating the analysis considerably, probably too much. 
       
\section{Summary}
Production of (pseudo-)scalar particles at LHC is a good way to probe gluon distributions, but 
unfortunately the effect of linear gluon polarization on Higgs production is small (2-5\% level), 
smaller than the current theoretical uncertainty in the perturbative treatment (NNLL+NNLO). 
$C=+$ quarkonium states may offer alternative probes, but in this case the predictions have large 
theoretical uncertainties. Future LHC data on bottomonium states $\chi_{b0/2}$ and $\eta_b$ are 
most promising. Linear gluon polarization is expected to lead to large differences between these 
three states. 

Heavy quark pair and dijet production in DIS at a high-energy EIC offer clean channels for probing 
linearly polarized gluons and the gluon Sivers effect.    
Specific $\cos 2\phi$ asymmetries may exhibit  large $h_1^{\perp\, g}$ effects, allowing to study its sign, 
its small-$x$ behavior. It may even show saturation effects, as it probes the WW or $f$-type ($[+,+]$) distribution, 
which is expected to show a significant change in behavior around $k_\perp \sim Q_s$. 
This same distribution happens to appear in Higgs or $J^{PC}=0^{\pm +}$ quarkonium production at LHC.
In contrast, for the gluon Sivers TMD the cleanest probes at EIC and at 
RHIC and/or AFTER@LHC are actually entirely complementary. 

\begin{acknowledgement}
I would like to thank Maarten Buffing, Wilco den Dunnen, Jean-Philippe Lansberg, Piet Mulders, Elena Petreska, Cristian Pisano, and Jian Zhou, for 
fruitful discussions and/or collaborations. 
\end{acknowledgement}


\end{document}